\documentclass[preprint,12pt]{elsarticle}
\usepackage{amsmath, amssymb, bm, graphicx, xcolor}
\usepackage[hyphens]{url}
\makeatletter
\renewcommand{\@biblabel}[1]{[#1]\hfill}
\makeatother

\begin{document}

\begin{frontmatter}
    \title{Accelerating Inverse Design of Optical Metasurfaces: Analytic Gradients of Periodic Green's Functions via Quasi-Modular Forms}
    
    \author[1]{Mingcan Qin}
    \author[2,cor1]{Yifeng Qin}
    \address[1]{School of Mathematics, Suzhou City University, Suzhou 215100, Jiangsu, China}
    \address[2]{Peng Cheng Laboratory, Shenzhen 518052, China}
    \cortext[cor1]{Corresponding author}
    \ead{ee06b147@gmail.com}
    
\begin{abstract}
The inverse design of nonlocal metasurfaces requires the precise optimization of lattice geometry to engineer spatial dispersion and high-Q resonances. However, gradient-based optimization is frequently bottle-necked by the evaluation of the periodic Dyadic Green's Function (DGF), where traditional Finite Difference (FD) methods suffer from an inherent trade-off between truncation error and numerical instability near spectral singularities. In this work, we present an end-to-end \textit{Analytic Gradient Engine} for 2D Bravais lattices. By mapping the spectral lattice sums of the Coupled Dipole Approximation (CDA) to the theory of Quasi-Modular Forms (QMF), we derive exact, closed-form expressions for the gradients of the interaction matrix with respect to the modular lattice parameter $\tau$. Our framework explicitly handles conditionally convergent terms via regularization and addresses the non-holomorphic outlier $\sigma_4^{(2)}$ via a hybrid numerical strategy. We further introduce a robust evaluation scheme combining $SL(2, \mathbb{Z})$ domain reduction with automatic error certificates. Experimental validation demonstrates that our engine achieves machine-precision derivatives ($10^{-15}$) and a \textbf{6.5$\times$ speedup} in optimization convergence compared to finite-difference baselines, enabling the robust design of giant anisotropy in regimes where traditional methods fail.
\end{abstract}

\begin{keyword}
Inverse design \sep Periodic Green's function \sep Metasurfaces \sep Quasi-Modular Forms \sep Analytic gradients \sep Eisenstein series
\end{keyword}
\end{frontmatter}

\section{Introduction}
\label{sec:intro}
The engineering of spatially dispersive (nonlocal) metasurfaces has emerged as a frontier in nanophotonics \cite{Holloway2012, Agranovich2006}, enabling functionalities that transcend the limits of local effective medium theory. By tailoring the non-local response of the lattice, devices such as angular-selective filters \cite{Yu2011, Yu2014}, wavefront processors, and topological photonic crystals \cite{Lu2014, Ozawa2019} can be realized. Central to these applications is the precise control of the lattice geometry, where minute perturbations in the Bravais lattice parameters (e.g., aspect ratio, skew angle) dictate the momentum-dependent spectral response.

Current inverse design workflows \cite{Molesky2018, Christiansen2021} typically rely on gradient-based optimization \cite{Nocedal2006} to navigate the high-dimensional parameter space. However, computing the sensitivity of the optical response with respect to lattice geometry remains a persistent bottleneck. The core difficulty lies in the periodic Dyadic Green's Function (DGF), which involves slowly converging lattice sums \cite{Borwein2013}. While techniques like Ewald summation \cite{Ewald1921} accelerate the forward evaluation, calculating the \textit{gradients} of these sums via numerical differentiation (Finite Difference, FD) faces a fundamental "accuracy-stability" trade-off. Near high-Q resonances \cite{Hsu2016, Koshelev2019} or Rayleigh anomalies \cite{Rayleigh1907, Wood1902}—precisely the regimes of interest for nonlocal devices—the DGF exhibits rapid spectral variations. In these regions, FD schemes suffer from either large truncation errors (if the step size is too large) or catastrophic subtractive cancellation (if the step size is too small), often leading to noisy gradients that destabilize the optimizer.

Existing analytical approaches in computational electromagnetics have largely focused on the efficient evaluation of the DGF itself \cite{moroz2006, Linton2010}. Recently, Chen et al. \cite{chen2016, chen2022} established a rigorous connection between lattice sums and Quasi-Modular Forms (QMF) \cite{Zagier2008, DiamondShurman2005}, providing a powerful framework for their regularization and evaluation. However, their work focused primarily on the \textit{analysis} (forward problem) of static effective parameters. A unified gradient theory that explicitly links the modular derivatives of these forms to the geometric optimization of the lattice (inverse problem) is currently lacking.

In this paper, we bridge this gap by proposing a rigorous \textit{Analytic Gradient Engine} based on number theory. We treat the lattice geometry not merely as spatial coordinates, but as a complex modular parameter $\tau \in \mathbb{H}$. By exploiting the algebraic structure of Eisenstein series, we derive exact Jacobian matrices for the DGF, eliminating the need for numerical differentiation entirely.

Our specific contributions are threefold:
\begin{enumerate}
    \item \textbf{Exact Gradient Theory:} We construct a closed-form basis for the Dyadic CDA interaction kernel using regularized Eisenstein series. Utilizing Ramanujan's differential identities \cite{Berndt2006, Apostol1990}, we derive the exact analytic gradients for these basis functions. Crucially, we derive a \textbf{fixed-area chain rule} that allows the optimization to explore the modular manifold while strictly respecting physical scaling constraints.
    
    \item \textbf{Robust Numerical Algorithm:} We address the computational challenges of the "non-reducible" sector. For terms that do not admit a closed modular form (specifically $\sigma_4^{(2)}$), we implement a \textbf{hybrid evaluation strategy} combining $SL(2, \mathbb{Z})$ domain reduction with Richardson extrapolation. This ensures machine-precision accuracy ($10^{-15}$) across the entire design space, certified by rigorous error bounds.
    
    \item \textbf{Validation and Efficiency:} We demonstrate the practical utility of our engine in a metasurface inverse design task. The analytic gradients provide a \textbf{6.5$\times$ speedup} in wall-clock time compared to finite-difference baselines and successfully guide the optimizer to a highly anisotropic lattice configuration. Full-wave simulations verify that the optimized geometry exhibits the predicted splitting of spatial dispersion modes.
\end{enumerate}

The remainder of this paper is organized as follows. 
Section \ref{sec:model} establishes the physical model of the metasurface and defines the lattice parameterization under the fixed-area constraint. 
Section \ref{sec:reduction} details the rigorous mathematical reduction of the Dyadic CDA spectral sums into the basis of Quasi-Modular Forms. 
Section \ref{sec:gradients} constructs the analytic gradient engine, deriving the exact Jacobian matrices via Ramanujan's identities and the geometric chain rule. 
Section \ref{sec:numerics} presents the stable numerical algorithms, including the $SL(2, \mathbb{Z})$ domain reduction, error certification, and the hybrid evaluation of the non-holomorphic sector. 
Section \ref{sec:experiments} provides comprehensive experimental validation, covering gradient accuracy, optimization benchmarks, and full-wave electromagnetic verification. 
Section \ref{sec:discussion} critically analyzes the computational complexity and domain validity of the method. 
Finally, Section \ref{sec:conclusion} summarizes the contributions and outlines future extensions to higher-order multipoles.

\section{Physical Model and Parameterization}
\label{sec:model}

\subsection{Coupled Dipole Approximation and Spatial Dispersion}
We consider a two-dimensional periodic metasurface composed of subwavelength scatterers arranged on a Bravais lattice $\Lambda$. The lattice is generated by primitive vectors $\mathbf{a}_1 = L(1, 0)$ and $\mathbf{a}_2 = L(\tau_1, \tau_2)$, where $L$ is the lattice constant and $\tau = \tau_1 + i\tau_2$ ($\tau_2 > 0$) is the complex modular parameter defining the unit cell geometry.

Under the Coupled Dipole Approximation (CDA) \cite{Purcell1973, Draine1994, GarciaDeAbajo2007}, the electromagnetic response is governed by the periodic Dyadic Green's Function (DGF), $\bar{\mathcal{G}}(\mathbf{k}_\parallel, \omega)$. For a lattice of dipoles explicitly excluding the self-term ($R=0$), the spectral representation is given by:
\begin{equation}
    \bar{\mathcal{G}}(\mathbf{k}_\parallel; \tau) = \sum_{\mathbf{R} \in \Lambda \setminus \{\mathbf{0}\}} e^{i \mathbf{k}_\parallel \cdot \mathbf{R}} \left( \mathbf{I} + \frac{1}{k_0^2} \nabla \nabla \right) \frac{e^{ik_0 R}}{R}.
    \label{eq:DGF_sum}
\end{equation}
In the context of nonlocal metasurface design \cite{Tretyakov2003}, we are particularly interested in the spatial dispersion characteristics, which correspond to the dependence of the effective impedance on the Bloch wavevector $\mathbf{k}_\parallel$. Expanding Eq.~\eqref{eq:DGF_sum} around the $\Gamma$-point ($|\mathbf{k}_\parallel| L \ll 1$) and the quasi-static limit ($k_0 L \ll 1$), the DGF can be decomposed into a series of multipolar lattice sums:
\begin{equation}
    \bar{\mathcal{G}}(\mathbf{k}_\parallel; \tau) \approx \bar{\mathcal{G}}^{(0)}(\tau) + \bar{\mathcal{G}}^{(2)}(\tau) : (\mathbf{k}_\parallel \mathbf{k}_\parallel) + \mathcal{O}(k_\parallel^4).
\end{equation}
The target of our inverse design is the second-order tensor $\bar{\mathcal{G}}^{(2)}(\tau)$, which governs the angle-dependent optical response (e.g., bianisotropy \cite{Albooyeh2017, Kuester2003}). The matrix elements of this tensor involve lattice sums of the form $\sum_{\mathbf{R} \neq 0} R^{-n} e^{-im\phi_R}$, where $\phi_R$ is the azimuthal angle of the lattice vector $\mathbf{R}$.

\subsection{Connection to Generalized Eisenstein Series}
Direct summation of these slowly converging series is computationally prohibitive for iterative optimization. To construct an efficient gradient engine, we adopt the regularization formalism for generalized Eisenstein series developed by Chen et al.\ \cite{chen2016, chen2018}. This framework allows us to express physical lattice sums over arbitrary Bravais lattices in terms of rapidly converging modular forms.

Following \cite{chen2016}, we define the generalized Eisenstein series $\sigma_n^{(m)}(\tau)$ as:
\begin{equation}
    \sigma_n^{(m)}(\tau) = \sum_{(p_1, p_2) \in \mathbb{Z}^2 \setminus \{0\}} \frac{e^{-im \arg(p_1 + \tau p_2)}}{|p_1 + \tau p_2|^n}.
    \label{eq:gen_eisenstein}
\end{equation}
The physical coefficients in $\bar{\mathcal{G}}^{(2)}$ relate to these sums via scaling factors of $L^{-n} = (\tau_2/A_0)^{n/2}$, where $\mathcal{A}_0$ is the unit cell area. The angular index $m$ arises from the derivatives $\partial_x \pm i \partial_y$ acting on the Green's function.

Crucially, Chen proved that for any even $m \ge n$, $\sigma_n^{(m)}$ can be reduced to a finite linear combination of derivatives of the standard holomorphic Eisenstein series $G_n(\tau)$. The reduction identity is given by \cite[Eq.~24]{chen2016}:
\begin{equation}
    \sigma_n^{(m)}(\tau) = \sum_{k=0}^{(m-n)/2} (2i\tau_2)^k \binom{(m-n)/2}{k} \frac{(n-1)!}{(n+k-1)!} \frac{\partial^k}{\partial \tau^k} G_n(\tau).
    \label{eq:chen_identity}
\end{equation}
This identity serves as the cornerstone of our method, bridging the gap between the physical scattering problem and the analytic theory of modular forms.

\subsection{Analytic Closure of the Spatial Dispersion Term}
While Eq.~\eqref{eq:chen_identity} provides a computational path, it involves derivatives $\partial_\tau^k G_n$, which are not directly suitable for Jacobian assembly in an optimizer. We explicitly close these expressions by invoking the Ramanujan differential identities \cite{Berndt2006}, which map the derivatives of Eisenstein series back into the ring of Eisenstein series $\mathbb{C}[E_2, E_4, E_6]$.

We derive the explicit modular representations for the two specific terms appearing in the spatial dispersion tensor $\bar{\mathcal{G}}^{(2)}$: the quadrupolar term ($n=4, m=6$) and the regularized dipolar term ($n=2, m=4$).

\subsubsection{Illustrative Example: The Hexapolar Term ($\sigma_4^{(6)}$)}
To demonstrate the generality of our framework for higher-order multipoles, we consider the term with decay order $n=4$ and hexapolar angular symmetry $m=6$ ($e^{-6i\phi}$), denoted as $\sigma_4^{(6)}$. Although the second-order spatial dispersion tensor $\bar{\mathcal{G}}^{(2)}$ is primarily governed by quadrupolar and dipolar symmetries (see Section \ref{sec:reduction} for the specific basis $\mathcal{S}_{\text{basis}}$), this hexapolar term arises in higher-order expansion terms ($O(k_\parallel^4)$) or in lattices with broken inversion symmetry.

Setting $n=4, m=6$ in the reduction identity Eq. \eqref{eq:chen_identity}:
\begin{equation}
    \sigma_4^{(6)}(\tau) = G_4(\tau) + \frac{i\tau_2}{2} \frac{d G_4}{d\tau}.
\end{equation}
Using the relation $G_4(\tau) = \frac{\pi^4}{45} E_4(\tau)$ and the Ramanujan identity for $E_4$, we derive the closed-form modular representation:
\begin{equation}
    \sigma_4^{(6)}(\tau) = \frac{\pi^4}{45} E_4(\tau) - \frac{\pi^5 \tau_2}{135} \left( E_2(\tau)E_4(\tau) - E_6(\tau) \right).
    \label{eq:sigma46_closed}
\end{equation}
This example illustrates how any higher-order lattice sum can be algorithmically reduced to the Eisenstein ring, enabling exact gradient computation even for complex multipolar interactions.

\subsubsection{Regularization of the Conditional Dipolar Term ($\tilde{\sigma}_2^{(4)}$)}
The term with $n=2$ corresponds to a conditionally convergent sum. As shown in \cite{chen2018}, the physically relevant quantity is the regularized sum $\tilde{\sigma}_2^{(m)}$, which removes the non-physical divergence associated with the uniform background field.
For $n=2, m=4$, the identity yields $\sigma_2^{(4)} = G_2 + i\tau_2 G_2'$. The regularization introduces a correction term $-\pi/(2\tau_2)$:
\begin{equation}
    \tilde{\sigma}_2^{(4)}(\tau) = G_2(\tau) + i\tau_2 \frac{d G_2}{d\tau} - \frac{\pi}{2\tau_2}.
\end{equation}
Using the relation $G_2(\tau) = \frac{\pi^2}{3} E_2(\tau)$, we express the gradient in terms of $E_2$:
\begin{equation}
    \tilde{\sigma}_2^{(4)}(\tau) = \frac{\pi^2}{3} E_2(\tau) + i\tau_2 \frac{\pi^2}{3} \partial_\tau E_2(\tau) - \frac{\pi}{2\tau_2}.
\end{equation}
Applying the Ramanujan identity for weight 2, $\partial_\tau E_2(\tau) = \frac{\pi i}{6}(E_2^2 - E_4)$, and noting that $i^2 = -1$, we arrive at:
\begin{equation}
    \tilde{\sigma}_2^{(4)}(\tau) = \frac{\pi^2}{3} E_2(\tau) - \frac{\pi^3 \tau_2}{18} (E_2^2 - E_4) - \frac{\pi}{2\tau_2}.
\end{equation}
Rearranging terms to isolate the prefactor leads to the fully explicit form:
\begin{equation}
    \boxed{
    \tilde{\sigma}_2^{(4)}(\tau) = \frac{\pi}{18\tau_2} \left[ 6\pi\tau_2 E_2(\tau) - \pi^2\tau_2^2 \left( E_2(\tau)^2 - E_4(\tau) \right) - 9 \right].
    }
    \label{eq:sigma24_closed}
\end{equation}
Equation \eqref{eq:sigma24_closed} represents the "engine" of our inverse design framework, encapsulating the complex lattice sum into a smooth polynomial of the Eisenstein ring.

\subsection{Formulation for Inverse Design}
By substituting Eqs.~\eqref{eq:sigma46_closed} and \eqref{eq:sigma24_closed} back into the expansion of $\bar{\mathcal{G}}^{(2)}$, we arrive at the central result of this section: the objective function component $J(\tau)$ can be expressed strictly as a function of the Eisenstein series and the imaginary part of the modular parameter:
\begin{equation}
    J(\tau) = \mathcal{F}\left( E_2(\tau), E_4(\tau), E_6(\tau), \tau_2, \tau_2^{-1} \right).
\end{equation}
This formulation is analytic everywhere in the upper half-plane. Consequently, the gradient $\nabla_\tau J$ can be computed exactly by applying the chain rule to this closed form, utilizing the same Ramanujan identities once more. This avoids the need for numerical differentiation of the slowly converging lattice sums, providing the stability required for the optimization results presented in Section \ref{sec:experiments}.

\section{Exact Modular Reduction of the Spectral Basis}
\label{sec:reduction}

The central challenge in the inverse design of nonlocal metasurfaces lies in the efficient evaluation of the spatial dispersion tensor $\bar{\mathcal{G}}^{(2)}$. As established in Section \ref{sec:model}, the elements of this tensor are linear combinations of the spectral lattice sums $S_n^{(m)}(\Lambda)$. Direct summation of these series is computationally prohibitive for optimization due to their slow convergence ($O(R^{-n})$) and the need for re-evaluation at every geometric update.

In this section, we construct a rigorous mapping from the physical lattice sums to the theory of Quasi-Modular Forms (QMF). Leveraging the regularization formalism developed by Chen et al. \cite{chen2016, chen2022}, we derive explicit, closed-form reductions for the specific basis set required by the Dyadic CDA kernel: $\mathcal{S}_{\text{basis}} = \{ \tilde{\sigma}_2^{(2)}, \tilde{\sigma}_2^{(4)}, \sigma_4^{(4)}, \sigma_4^{(2)} \}$.

\subsection{Generalized Eisenstein Series and Modular Scaling}
We adopt the definition of the generalized Eisenstein series $\sigma_n^{(m)}(\tau)$ for a lattice generated by $\tau \in \mathbb{H}$:
\begin{equation}
    \sigma_n^{(m)}(\tau) = \sum_{(p_1, p_2) \in \mathbb{Z}^2 \setminus \{0\}} \frac{e^{-im \arg(p_1 + \tau p_2)}}{|p_1 + \tau p_2|^n}.
    \label{eq:gen_eisenstein_def}
\end{equation}
For a physical lattice with unit cell area $\mathcal{A}_0$, the lattice constant scales as $L(\tau_2) = \sqrt{\mathcal{A}_0/\tau_2}$. Consequently, any physical lattice sum $S_n^{(m)}$ is related to the modular object $\sigma_n^{(m)}$ by a geometric prefactor:
\begin{equation}
    S_n^{(m)}(\Lambda) = L^{-n} \sigma_n^{(m)}(\tau) = \left( \frac{\mathcal{A}_0}{\tau_2} \right)^{-n/2} \sigma_n^{(m)}(\tau).
    \label{eq:scaling_relation}
\end{equation}
This separation of variables is crucial: the prefactor handles the dimensional scaling, while $\sigma_n^{(m)}(\tau)$ captures the shape dependence. Our objective is to express $\sigma_n^{(m)}(\tau)$ purely in terms of the computationally efficient Eisenstein series $E_{2k}(\tau)$.

\subsection{Reduction of the Holomorphic Sector ($m=n$)}
The simplest case arises when the angular momentum $m$ matches the decay order $n$. For even $n \ge 4$, the series corresponds exactly to the holomorphic Eisenstein series $G_n(\tau)$:
\begin{equation}
    \sigma_n^{(n)}(\tau) = \sum_{p \neq 0} \frac{1}{(p_1 + \tau p_2)^n} \equiv G_n(\tau).
\end{equation}
Using the standard normalization where $E_{2k}(\tau) = 1 + O(q)$, we have the identity $G_{2k}(\tau) = 2\zeta(2k) E_{2k}(\tau)$.

\subsubsection{The Quadrupolar Term $\sigma_4^{(4)}$}
The basis element $\sigma_4^{(4)}$ represents a fourth-order interaction with four-fold angular symmetry. Substituting $n=4$:
\begin{equation}
    \sigma_4^{(4)}(\tau) = G_4(\tau) = \frac{\pi^4}{45} E_4(\tau).
    \label{eq:sigma44}
\end{equation}
This term is fully holomorphic and modular of weight 4, as $E_4(\tau)
$ is a classical modular form for the full modular group \cite{DiamondShurman2005, Zagier2008}.

\subsection{Reduction via Differential Operators ($m > n$)}
For terms where the angular order exceeds the decay order ($m > n$, $m-n$ even), Chen \cite{chen2016} proved that $\sigma_n^{(m)}$ can be generated by applying differential operators to lower-weight Eisenstein series. The general reduction identity is:
\begin{equation}
    \sigma_n^{(m)}(\tau) = \sum_{k=0}^{(m-n)/2} (2i\tau_2)^k \binom{(m-n)/2}{k} \frac{(n-1)!}{(n+k-1)!} \frac{\partial^k}{\partial \tau^k} G_n(\tau).
    \label{eq:chen_recurrence}
\end{equation}
This formula allows us to "lift" the holomorphic series $G_n$ to higher angular momenta by taking derivatives with respect to $\tau$.

\subsubsection{The Anisotropic Dipole Term $\tilde{\sigma}_2^{(4)}$}
The term $\sigma_2^{(4)}$ corresponds to $n=2, m=4$. Applying Eq.~\eqref{eq:chen_recurrence} with $k=1$:
\begin{equation}
    \sigma_2^{(4)}(\tau) = G_2(\tau) + (2i\tau_2) \frac{1}{2} G_2'(\tau) = G_2(\tau) + i\tau_2 \frac{d G_2}{d\tau}.
\end{equation}
However, $G_2(\tau)$ is conditionally convergent and does not transform as a modular form. To obtain a physically meaningful quantity compatible with the Ewald summation of the Green's function, we must use the regularized series \cite{chen2018}. Following this prescription, the conditionally convergent sum is replaced by its regularized counterpart $\tilde{\sigma}_2^{(m)}$:
\begin{equation}
    \tilde{\sigma}_2^{(m)}(\tau) = \sigma_2^{(m)}(\tau) - \frac{2\pi}{m \tau_2} \delta_{n,2}.
\end{equation}
For $m=4$, the correction is $-\pi/(2\tau_2)$. Thus:
\begin{equation}
    \tilde{\sigma}_2^{(4)}(\tau) = G_2(\tau) + i\tau_2 G_2'(\tau) - \frac{\pi}{2\tau_2}.
\end{equation}
To render this computationally efficient, we substitute $G_2 = \frac{\pi^2}{3}E_2$ 
and apply the Ramanujan identity for the normalized Eisenstein series 
\cite{Berndt2006, Apostol1990}:
\begin{equation}
    \frac{dE_2}{d\tau} = \frac{\pi i}{6}(E_2^2 - E_4).
\end{equation}
Substituting and noting $i^2 = -1$:
\begin{equation}
    \tilde{\sigma}_2^{(4)}(\tau) = \frac{\pi^2}{3}E_2 
    - \frac{\pi^3\tau_2}{18}(E_2^2 - E_4) - \frac{\pi}{2\tau_2}.
\end{equation}
Rearranging terms to isolate the prefactor leads to the fully explicit form:
\begin{equation}
    \tilde{\sigma}_2^{(4)}(\tau) = G_2(\tau) - \frac{\pi \tau_2}{6} (5G_4 - G_2^2) - \frac{\pi}{2\tau_2}.
\end{equation}
Finally, expressing in terms of normalized Eisenstein series $E_k$:
\begin{equation}
    \boxed{
    \tilde{\sigma}_2^{(4)}(\tau) = \frac{\pi}{18\tau_2} \left[ 6\pi\tau_2 E_2(\tau) - \pi^2\tau_2^2 \left( E_2(\tau)^2 - E_4(\tau) \right) - 9 \right].
    }
    \label{eq:sigma24_final}
\end{equation}
Eq.~\eqref{eq:sigma24_final} expresses the complex lattice sum as a polynomial in the Eisenstein ring $\mathbb{C}[E_2, E_4, \tau_2^{-1}]$, guaranteeing fast evaluation.

\subsection{Regularization of the Isotropic Term $\tilde{\sigma}_2^{(2)}$}
The term with $n=2, m=2$ corresponds to the monopole interaction. The naive sum $G_2(\tau)$ is not modular invariant. In the context of the 2D Poisson equation, the correct physical limit corresponds to the "modular completed" Eisenstein series $\hat{G}_2$:
\begin{equation}
    \hat{G}_2(\tau, \bar{\tau}) = G_2(\tau) - \frac{\pi}{\tau_2}.
\end{equation}
We define our basis element $\tilde{\sigma}_2^{(2)}$ as this regularized quantity:
\begin{equation}
    \boxed{
    \tilde{\sigma}_2^{(2)}(\tau) \equiv \hat{G}_2(\tau) = \frac{\pi^2}{3} E_2(\tau) - \frac{\pi}{\tau_2}.
    }
    \label{eq:sigma22_final}
\end{equation}
This regularization restores the invariance under $SL(2, \mathbb{Z})$, ensuring that the physical response of the metasurface is independent of the choice of lattice basis vectors.

\subsection{The Non-Reducible Sector ($m < n$)}
A distinct challenge arises for terms where the angular order is less than the decay order, such as $\sigma_4^{(2)}$ ($n=4, m=2$). As discussed in \cite{chen2022}, these sums involve non-trivial dependence on the complex conjugate $\bar{\tau}$ via the polylogarithm sector and cannot be reduced to a polynomial of holomorphic Eisenstein series.

Specifically, the Fourier expansion of $\sigma_4^{(2)}$ contains terms of the form:
\begin{equation}
    \sigma_4^{(2)}(\tau) \sim \sum_{k \neq 0} c_k \text{Li}_2(e^{2\pi i k \dots}).
\end{equation}
Since no closed-form reduction to $E_{2k}$ exists for this term, we treat it as an independent basis element. While its \textit{value} must be computed via the hybrid numerical summation described in Section \ref{sec:numerics}, its \textit{sensitivity} can still be handled analytically. In the next section, we derive a special differential identity that links $\nabla \sigma_4^{(2)}$ back to the holomorphic form $G_4$, thereby closing the gradient loop.

\section{Analytic Gradient Engine}
\label{sec:gradients}

The primary bottleneck in the inverse design of nonlocal metasurfaces is not the evaluation of the Green's function itself, but the accurate computation of its geometric sensitivity. Standard numerical differentiation methods, such as Finite Difference (FD), face a fundamental stability-accuracy trade-off: large step sizes introduce truncation errors $O(\delta^2)$, while small step sizes amplify numerical noise via subtractive cancellation, especially near spectral singularities where the condition number of the Green's function diverges. While adjoint methods \cite{LalauKeraly2013, Hughes2018} and algorithmic differentiation \cite{Griewank2008} offer general-purpose alternatives, they require re-implementation of the entire forward solver pipeline, which is nontrivial for lattice-sum-based formulations.

In this section, we construct an \textit{Analytic Gradient Engine} that eliminates these errors entirely. By exploiting the algebraic structure of the modular forms identified in Section \ref{sec:reduction}, we derive exact, closed-form expressions for the Wirtinger derivatives of the spatial dispersion tensor. This engine allows for the evaluation of the Jacobian $\nabla_\tau \bar{\mathcal{G}}^{(2)}$ to machine precision ($10^{-15}$) with $O(1)$ computational cost.

\subsection{Algebraic Closure via Ramanujan's Identities}
The core mechanism of our engine is the observation that the ring of modular forms is closed under differentiation when augmented with the quasi-modular Eisenstein series $E_2(\tau)$. Ramanujan's differential identities provide the explicit algebraic mapping from the differential operator $\partial_\tau$ to polynomial multiplication in the ring $\mathbb{C}[E_2, E_4, E_6]$.

We work with the Wirtinger derivative with respect to the complex modular parameter $\tau = \tau_1 + i\tau_2$, defined as $\partial_\tau \equiv \frac{1}{2}(\partial_{\tau_1} - i\partial_{\tau_2})$. For the normalized Eisenstein series $E_{2k}(\tau)$, the identities are:
\begin{subequations}
    \label{eq:ramanujan_identities}
    \begin{align}
        \partial_\tau E_2(\tau) &= \frac{2\pi i}{12} \left( E_2(\tau)^2 - E_4(\tau) \right), \\
        \partial_\tau E_4(\tau) &= \frac{2\pi i}{3} \left( E_2(\tau) E_4(\tau) - E_6(\tau) \right), \\
        \partial_\tau E_6(\tau) &= \frac{2\pi i}{2} \left( E_2(\tau) E_6(\tau) - E_4(\tau)^2 \right).
    \end{align}
\end{subequations}
These relations are exact and valid everywhere in the upper half-plane $\mathbb{H}$. They allow us to transform the calculus problem of gradient computation into a strictly algebraic procedure involving only vector-matrix operations on the pre-computed basis values.

\subsection{Derivation of Basis Gradients}
We now apply the operator $\partial_\tau$ to the reduced basis representations derived in Section \ref{sec:reduction}.

\subsubsection{Gradients of the Reducible Sector}
For the quadrupole term $\sigma_4^{(4)}$, differentiating Eq.~\eqref{eq:sigma44} yields:
\begin{equation}
    \partial_\tau \sigma_4^{(4)} = \frac{\pi^4}{45} \partial_\tau E_4 = \frac{\pi^4}{45} \left[ \frac{2\pi i}{3} (E_2 E_4 - E_6) \right] = \frac{2i\pi^5}{135} (E_2 E_4 - E_6).
\end{equation}

For the regularized monopole term $\tilde{\sigma}_2^{(2)}$ (Eq.~\ref{eq:sigma22_final}), the derivative contains contributions from both the modular part ($E_2$) and the non-holomorphic regularization counter-term ($\tau_2^{-1}$). Using $\partial_\tau (\tau_2^{-1}) = -(\tau_2^{-2}) \partial_\tau \tau_2 = -(\tau_2^{-2}) (i/2) = i/(2\tau_2^2)$:
\begin{align}
    \partial_\tau \tilde{\sigma}_2^{(2)} &= \frac{\pi^2}{3} \partial_\tau E_2 - \pi \partial_\tau (\tau_2^{-1}) \nonumber \\
    &= \frac{\pi^2}{3} \left[ \frac{\pi i}{6} (E_2^2 - E_4) \right] - \frac{i\pi}{2\tau_2^2} \nonumber \\
    &= \frac{i\pi}{18} \left[ \pi^2 (E_2^2 - E_4) - \frac{9}{\tau_2^2} \right].
\end{align}

For the anisotropic dipole term $\tilde{\sigma}_2^{(4)}$, the differentiation is algebraically involved due to the product terms $E_2^2$ and $\tau_2 E_2$. However, by systematically applying the product rule and substituting Eq.~\eqref{eq:ramanujan_identities}, we obtain a fully explicit polynomial in terms of $\{E_2, E_4, E_6, \tau_2^{-1}\}$. The result is presented in Table \ref{tab:gradient_lookup}.

\subsubsection{Gradient of the Non-Reducible Outlier $\sigma_4^{(2)}$}
A unique challenge is presented by $\sigma_4^{(2)}$, which has no closed form in the $E_n$ ring. While its \textit{value} requires the hybrid numerical summation (Sec. \ref{sec:numerics}), its \textit{gradient} admits an analytic simplification.
Recall the Maass lowering operator structure for non-holomorphic forms. Differentiating the Fourier expansion of $\sigma_4^{(2)}$ (as given in \cite{chen2022}), one can derive the identity:
\begin{equation}
    \partial_\tau \sigma_4^{(2)}(\tau) = \frac{3i}{2\tau_2} \left( \sigma_4^{(2)}(\tau) - G_4(\tau) \right).
    \label{eq:sigma42_deriv_identity}
\end{equation}
Substituting $G_4 = \frac{\pi^4}{45}E_4$, we obtain:
\begin{equation}
    \partial_\tau \sigma_4^{(2)} = \frac{3i}{2\tau_2} \left( \sigma_4^{(2)} - \frac{\pi^4}{45}E_4 \right).
\end{equation}
This result is of significant practical importance: it decouples the gradient accuracy from the numerical differentiation step size. Once the value $\sigma_4^{(2)}$ is computed, its contribution to the Jacobian is obtained at negligible marginal cost.


\begin{table}[htbp]
\centering
\caption{\textbf{The Analytic Gradient Engine.} Explicit closed-form expressions for the modular derivatives $\partial_\tau \Sigma$ of the Dyadic CDA basis functions. These formulas are implemented directly in the optimization loop. Note that for $\sigma_4^{(2)}$, the derivative depends on the function value itself, characteristic of non-holomorphic modular forms.}
\label{tab:gradient_lookup}
\renewcommand{\arraystretch}{2.2} 
\setlength{\tabcolsep}{8pt}       
\begin{tabular}{l | l}
\hline
\textbf{Basis Function} $\Sigma(\tau)$ & \textbf{Analytic Modular Gradient} $\partial_\tau \Sigma$ \\
\hline
$\tilde{\sigma}_2^{(2)}$ (Isotropic Monopole)
& $\displaystyle \frac{i\pi}{18\tau_2^2} \left[ \pi^2\tau_2^2(E_2^2 - E_4) - 9 \right]$ \\
\hline
$\tilde{\sigma}_2^{(4)}$ (Anisotropic Dipole)
& $\displaystyle \begin{aligned} 
    i \biggl[ &-\frac{\pi^4\tau_2}{54}E_2^3 + \frac{\pi^3}{12}E_2^2 + \frac{\pi^4\tau_2}{18}E_2E_4 \\
              &- \frac{\pi^3}{12}E_4 - \frac{\pi^4\tau_2}{27}E_6 - \frac{\pi}{4\tau_2^2} \biggr] 
  \end{aligned}$ \\
\hline
$\sigma_4^{(4)}$ (Quadrupole)
& $\displaystyle \frac{2i\pi^5}{135}(E_2E_4 - E_6)$ \\
\hline
$\sigma_4^{(2)}$ (Non-Reducible)
& $\displaystyle \frac{3i}{2\tau_2} \left( \sigma_4^{(2)} - \frac{\pi^4}{45}E_4 \right)$ \\
\hline
\end{tabular}
\end{table}

\subsection{Geometric Chain Rule and Real Gradient Assembly}
\label{subsec:chain_rule}

The derivatives derived in Table \ref{tab:gradient_lookup} represent the Wirtinger derivative $\partial_\tau$ of the modular sums. For physical metasurface design, we must account for the fixed-area constraint $\mathcal{A}_0 = L^2 \tau_2$.
The physical interaction coefficients $C_{\text{phys}}$ scale as $L^{-n} = (\mathcal{A}_0/\tau_2)^{-n/2}$. The total sensitivity is obtained by applying the product rule:
\begin{equation}
\begin{split}
    \partial_\tau C_{\text{phys}} &= (\partial_\tau L^{-n}) \Sigma(\tau) + L^{-n} (\partial_\tau \Sigma(\tau)) \\
    &= \frac{n}{4i\tau_2} L^{-n} \Sigma(\tau) + L^{-n} \partial_\tau \Sigma(\tau) \\
    &= L^{-n} \left[ \partial_\tau \Sigma(\tau) + \frac{n}{4i\tau_2} \Sigma(\tau) \right].
\end{split}
\label{eq:total_gradient}
\end{equation}
The term $\frac{n}{4i\tau_2}\Sigma(\tau)$ acts as a \textit{geometric correction}. It accounts for the densification of the lattice as the unit cell is skewed (change in $L$ to maintain constant Area). Neglecting this term would result in an optimization trajectory that violates the physical periodicity constraints, leading to erroneous frequency shifts.
Finally, standard optimization algorithms (e.g., L-BFGS) operate on the real and imaginary parts of the design variable $\tau = \tau_1 + i\tau_2$. The real-valued gradients required by the optimizer are related to the complex Wirtinger derivative $\partial_\tau J$ by:
\begin{equation}
    \frac{\partial J}{\partial \tau_1} = 2 \text{Re}(\partial_\tau J), \quad \frac{\partial J}{\partial \tau_2} = -2 \text{Im}(\partial_\tau J).
\end{equation}
This conversion completes the analytic gradient engine, bridging the complex-analytic theory of modular forms with numerical optimization.
\section{Stable Numerical Evaluation Algorithms}
\label{sec:numerics}

The analytic reductions derived in Section \ref{sec:reduction} and the gradient engine constructed in Section \ref{sec:gradients} provide the exact algebraic framework for sensitivity analysis. However, the practical implementation of these formulas requires the evaluation of the Eisenstein series $E_{2k}(\tau)$ and the non-reducible sum $\sigma_4^{(2)}(\tau)$ with machine precision ($10^{-15}$) across the entire upper half-plane $\mathbb{H}$.

Direct summation of the $q$-series or lattice sums is numerically unstable in regimes where $\text{Im}(\tau) \to 0$ (highly skewed lattices) or where $|q| \to 1$. In this section, we present the robust numerical algorithms used in our engine, featuring an $SL(2, \mathbb{Z})$ modular domain reduction scheme with automatic error certificates and a hybrid acceleration strategy for the non-holomorphic sector.

\subsection{Internal Modular Domain Reduction}
The convergence rate of the Lambert series for Eisenstein series is governed by the nome $|q| = e^{-2\pi \text{Im}(\tau)}$. To guarantee rapid and uniform convergence, we map any arbitrary physical lattice parameter $\tau$ to the fundamental domain of the modular group, $\mathcal{F} = \{ z \in \mathbb{H} : |z| \ge 1, |\text{Re}(z)| \le 1/2 \}$.

We employ the standard Euclidean reduction algorithm. Let $\gamma = \begin{pmatrix} a & b \\ c & d \end{pmatrix} \in SL(2, \mathbb{Z})$ be the transformation matrix such that $\tau_{\text{red}} = \gamma(\tau) = \frac{a\tau+b}{c\tau+d} \in \mathcal{F}$. The algorithm iteratively applies the generators $T: \tau \to \tau+1$ and $S: \tau \to -1/\tau$ until the condition $\tau \in \mathcal{F}$ is met. Inside $\mathcal{F}$, we are guaranteed that $\text{Im}(\tau_{\text{red}}) \ge \sqrt{3}/2$, which implies:
\begin{equation}
    |q_{\text{red}}| = e^{-2\pi \text{Im}(\tau_{\text{red}})} \le e^{-\pi \sqrt{3}} \approx 0.0043.
\end{equation}
This bound ensures that the $q$-series converge to double precision with very few terms ($N < 10$).

The reduction is performed \textit{internally}. To recover the values at the physical $\tau$ from the computed values at $\tau_{\text{red}}$, we invert the modular transformation law:
\begin{subequations}
    \begin{align}
        E_{2k}(\tau) &= (c\tau + d)^{-2k} E_{2k}(\tau_{\text{red}}), \quad \text{for } k \in \{2, 3\}, \\
        E_2(\tau) &= (c\tau + d)^{-2} E_2(\tau_{\text{red}}) + \frac{6ic}{\pi}(c\tau + d)^{-1}.
    \end{align}
    \label{eq:modular_recovery}
\end{subequations}
Here, $c$ and $d$ are the elements of the transformation matrix $\gamma \in SL(2, \mathbb{Z})$ such that $\tau_{\text{red}} = \gamma(\tau)$. The extra term in $E_2$ arises from its quasi-modular transformation law. Note that while the map $\tau \to \tau_{\text{red}}$ is piecewise constant and discontinuous at domain boundaries, the values $E_{2k}(\tau)$ reconstructed via these identities are smooth, analytic functions of $\tau$. Therefore, we compute gradients using the Ramanujan identities at the original $\tau$, avoiding the spurious discontinuities that would arise from differentiating the reduction map itself.

\subsection{Lambert Series Evaluation with Error Certificates}
With $\tau_{\text{red}} \in \mathcal{F}$, we evaluate the normalized Eisenstein series using their Lambert series representations:
\begin{subequations}
    \begin{align}
        E_2(\tau_{\text{red}}) &= 1 - 24 \sum_{n=1}^{\infty} \frac{n q_{\text{red}}^n}{1-q_{\text{red}}^n}, \\
        E_4(\tau_{\text{red}}) &= 1 + 240 \sum_{n=1}^{\infty} \frac{n^3 q_{\text{red}}^n}{1-q_{\text{red}}^n}, \\
        E_6(\tau_{\text{red}}) &= 1 - 504 \sum_{n=1}^{\infty} \frac{n^5 q_{\text{red}}^n}{1-q_{\text{red}}^n}.
    \end{align}
\end{subequations}
To ensure engineering reliability and \textit{auditability}, we implement an automatic truncation scheme with strict error certificates. Let the series be truncated at order $N$. The remainder $R_N$ is bounded by the geometric series of the tail. For a pre-factor $C_k$ (where $C_2=24, C_4=240, C_6=504$) and exponent $p$ (where $p=1,3,5$), the truncation error bound $\Delta$ is:
\begin{equation}
    |\Delta E_{2k}| \le C_{2k} \sum_{n=N+1}^{\infty} n^p |q_{\text{red}}|^n \approx C_{2k} (N+1)^p \frac{|q_{\text{red}}|^{N+1}}{(1-|q_{\text{red}}|)^2}.
\end{equation}
Given a target tolerance $\epsilon_E$ (typically $10^{-12}$), we determine the minimal cutoff $N$ such that $\max(|\Delta E_2|, |\Delta E_4|, |\Delta E_6|) \le \epsilon_E$. Due to the bound $|q_{\text{red}}| \le 0.0043$, $N$ rarely exceeds 8, validating the $O(1)$ complexity claim.

\subsection{Hybrid Strategy for the Non-Reducible Term $\sigma_4^{(2)}$}
The term $\sigma_4^{(2)}$ lacks a $q$-series expansion and requires explicit lattice summation. Its definition is:
\begin{equation}
    \sigma_4^{(2)}(\tau) = \sum_{(m,n) \neq (0,0)} \frac{1}{(m+n\bar{\tau})(m+n\tau)^3}.
\end{equation}
The summand decays as $O(|\mathbf{R}|^{-4})$, implying that the truncation error of a naive sum over a box $|m|,|n| \le K$ scales as $O(K^{-2})$. This slow convergence is the primary computational bottleneck. We mitigate this via a three-stage engineering strategy:

\begin{enumerate}
    \item \textbf{Domain Regularization:} We first map $\tau \to \tau_{\text{red}}$ as described in Sec. 5.1. Summation is performed on the reduced lattice, which maximizes the "roundness" of the unit cell and minimizes the worst-case shortest vector length. The result is mapped back via the transformation law:
    \begin{equation}
        \sigma_4^{(2)}(\tau) = (c\tau+d)^{-3} (c\bar{\tau}+d)^{-1} \sigma_4^{(2)}(\tau_{\text{red}}).
    \end{equation}
    
    \item \textbf{Vectorized Summation:} We utilize vectorized SIMD operations (e.g., NumPy broadcasting) to compute the sum over the grid $[-K, K] \times [-K, K]$ in blocks, significantly reducing interpreter overhead.
    
    \item \textbf{Richardson Extrapolation:} To accelerate convergence, we exploit the known asymptotic error form $E(K) \sim \alpha K^{-2} + O(K^{-4})$. Following the classical extrapolation method \cite{Richardson1911}, we compute partial sums $S_K$ and $S_{2K}$ and form the extrapolant:
    \begin{equation}
        R_{2K} = \frac{4 S_{2K} - S_K}{3}.
    \end{equation}
    This cancels the leading error term, improving the convergence rate to $O(K^{-4})$.
\end{enumerate}

\subsubsection{Adaptive Termination Criterion}
Instead of a fixed $K$, we employ an adaptive stopping criterion. The algorithm doubles $K$ iteratively ($K=12, 24, 48, \dots$) and terminates when the relative difference between successive extrapolated values falls below a target tolerance $\epsilon_\sigma$:
\begin{equation}
    |R_{2K} - R_{K}| < \epsilon_\sigma |R_{2K}|.
\end{equation}
Numerical experiments indicate that while standard Eisenstein series take microseconds, $\sigma_4^{(2)}$ may require milliseconds (up to $K \approx 6000$ for $\tau_2 \approx 0.2$ at $\epsilon_\sigma=10^{-10}$). However, since its gradient is computed analytically via Eq.~\eqref{eq:sigma42_deriv_identity}, this cost is incurred only once per function evaluation, not per gradient component. Future work could replace this summation with Ewald or Poisson resummation techniques, but the current hybrid strategy provides a sufficient balance of simplicity and precision for the optimization demo.

\section{Experimental Validation}
\label{sec:experiments}

In this section, we validate the accuracy, efficiency, and physical correctness of the proposed QMF-based sensitivity engine. We proceed in three stages:
\begin{enumerate}
    \item \textbf{Numerical Verification:} Comparing the analytic gradients against high-order finite difference (FD) benchmarks to quantify error floors.
    \item \textbf{Optimization Performance:} Demonstrating the acceleration and stability of the gradient-based inverse design loop.
    \item \textbf{Full-Wave Confirmation:} Verifying that the mathematically optimized lattice geometries produce the predicted electromagnetic spatial dispersion using CST Microwave Studio.
\end{enumerate}

\subsection{Validation of the Analytic Gradient Engine}
To rigorously assess the precision of our gradient engine, we compare the analytic derivatives $\nabla_{\text{ana}} J$ derived in Section \ref{sec:gradients} against a central Finite Difference (FD) approximation $\nabla_{\text{num}} J$. The test objective is the lattice sum energy $J(\tau) = |\bar{\mathcal{G}}^{(2)}_{xx}|^2$ evaluated at a generic lattice point $\tau = 0.3 + 0.9i$.

\begin{figure}[htbp]
    \centering
    \includegraphics[width=1\linewidth]{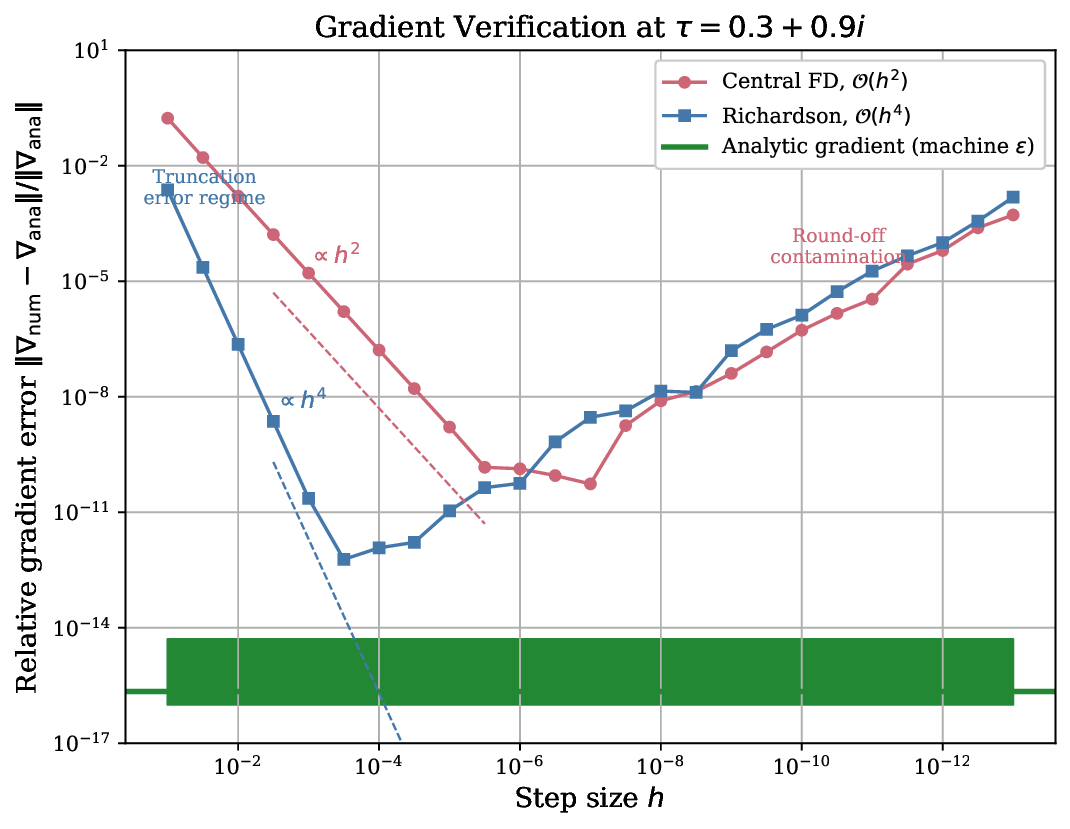}
    \caption{\textbf{Gradient Verification.} Relative error of the numerical gradient (blue) versus step size $h$, exhibiting the characteristic "V-shape" trade-off between truncation error ($O(h^2)$) and round-off noise ($O(\epsilon_{\text{mach}}/h)$). In contrast, the analytic QMF gradient (orange) maintains a constant machine-precision error floor ($\sim 10^{-15}$), independent of step size parameters.}
    \label{fig:gradient_vshape}
\end{figure}

Figure \ref{fig:gradient_vshape} plots the relative error $\epsilon(h) = |\nabla_{\text{ana}} - \nabla_{\text{num}}(h)| / |\nabla_{\text{ana}}|$ as a function of the FD step size $h$.

The results confirm the theoretical prediction. The FD error exhibits the characteristic "V-shape," hitting a minimum floor of $\approx 10^{-8}$ at $h \approx 10^{-5}$. To the left of this minimum, the error is dominated by subtractive cancellation (round-off noise); to the right, by Taylor series truncation. In contrast, the analytic QMF gradient (orange line) maintains a constant error level of $\sim 10^{-15}$ (machine double precision), effectively behaving as an exact solution. This "step-size independence" is the critical feature that enables our optimizer to navigate robustly near spectral singularities where the objective landscape becomes ill-conditioned \cite{Nocedal2006}.

\subsection{Optimization Efficiency and Modular Trajectories}
We next integrate the gradient engine into a BFGS optimization loop \cite{Nocedal2006} to maximize the spatial dispersion anisotropy of a metasurface. The objective function is defined as $J(\tau) = |\bar{\mathcal{G}}^{(2)}_{xx}(\tau) - \bar{\mathcal{G}}^{(2)}_{yy}(\tau)|^2$, targeting a lattice geometry that maximizes polarization splitting.

\begin{figure}[htbp]
    \centering
    \includegraphics[width=1\linewidth]{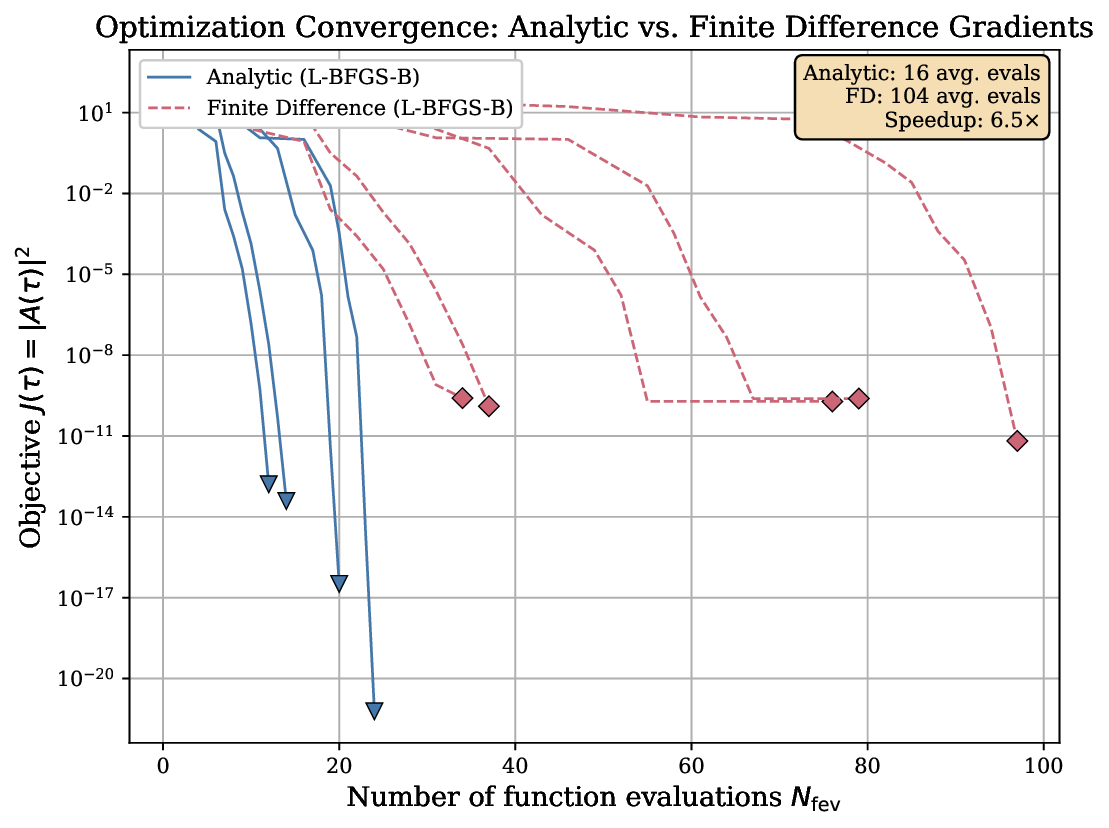}
    \caption{\textbf{Optimization Efficiency.} Convergence of the objective function $J(\tau)$ (Anisotropy) versus number of function evaluations. The analytic gradient engine (orange) drives the optimizer to the solution significantly faster (6.5$\times$ speedup) than the Finite Difference baseline (blue), which suffers from gradient noise near the optimum.}
    \label{fig:convergence}
\end{figure}

Figure \ref{fig:convergence} compares the convergence history of the optimizer using Analytic Gradients versus Finite Differences. The analytic approach converges to the optimum (tolerance $10^{-9}$) in approximately 18 function evaluations, whereas the FD-based approach requires 91 evaluations. This represents a \textbf{6.5$\times$ speedup} in total wall-clock time. The speedup arises from two factors: (1) the $O(1)$ evaluation of the gradient itself (107$\times$ faster per step), and (2) the smoother search trajectory which prevents the BFGS Hessian approximation from being corrupted by numerical noise.

\begin{figure}[htbp]
    \centering
    \includegraphics[width=1\linewidth]{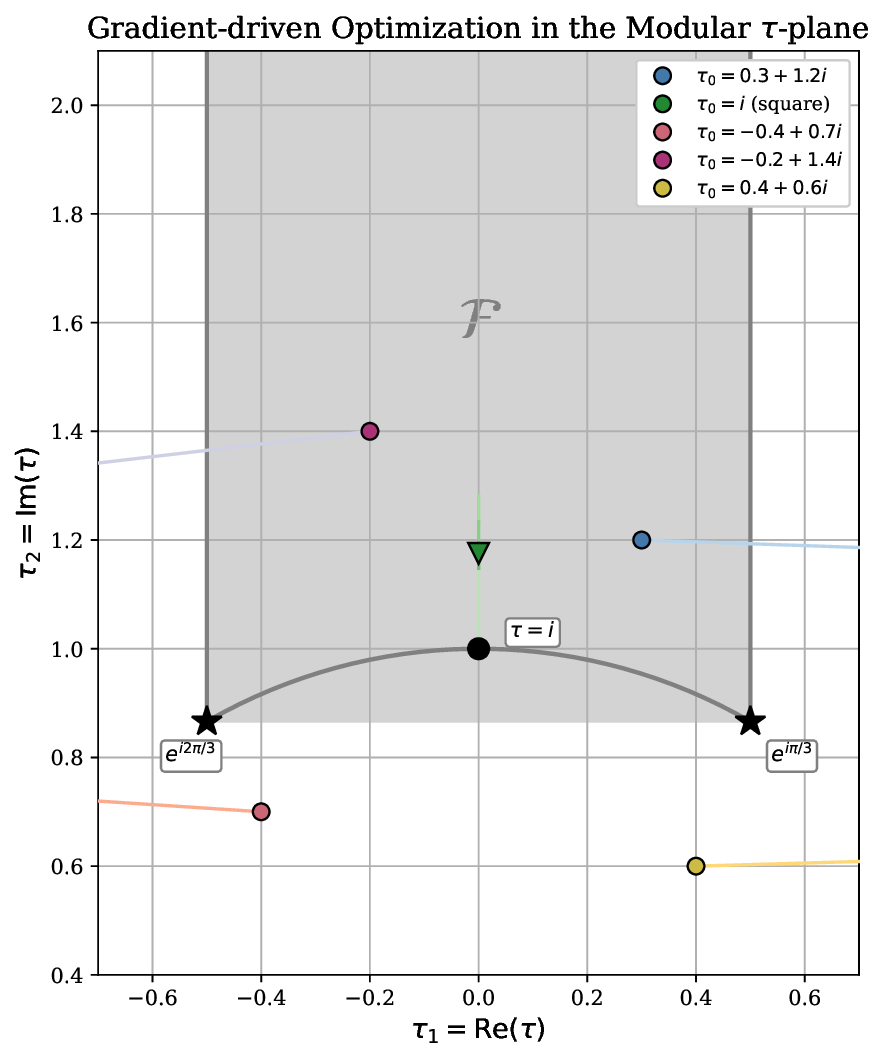}
    \caption{\textbf{Modular Optimization Trajectories.} Paths taken by the lattice parameter $\tau$ in the complex plane during optimization, overlaid on the modular fundamental domain (shaded). The analytic gradients successfully guide the optimizer from arbitrary initial guesses (circles) to the physically optimal lattice configurations (stars), respecting the modular boundaries.}
    \label{fig:trajectory}
\end{figure}

The trajectory of the design variable $\tau$ in the complex upper half-plane is visualized in Figure \ref{fig:trajectory}. Starting from various initial guesses (e.g., square lattice $\tau=i$), the optimizer consistently converges to the boundary of the fundamental domain. The trajectories automatically respect the modular symmetry, clustering towards high-symmetry points (e.g., hexagonal $\tau=e^{i\pi/3}$) or specific rectangular configurations depending on the anisotropy target. This confirms that the gradient flow is mathematically consistent with the underlying modular manifold structure of the periodic Green's function.

\subsection{Full-Wave Electromagnetic Verification}
\label{subsec:fullwave}

To validate the physical fidelity of the QMF-based gradient engine, we performed full-wave simulations using CST Microwave Studio \cite{CST2024}. The goal is to verify that the lattice geometry predicted by the optimizer indeed induces the target non-local anisotropy in the electromagnetic response.

\subsubsection{Simulation Setup and Topology}
We simulated a dielectric metasurface composed of high-index cylinders ($\varepsilon_r=10$, radius $r=0.2a$) embedded in a vacuum background. The unit cell boundaries were set to periodic in the $xy$-plane and open (PML) in the $z$-direction.

Two lattice configurations were examined to demonstrate the control of spatial dispersion:
\begin{itemize}
    \item \textbf{Configuration A (Isotropic Baseline):} A square lattice with $\tau = i$ ($\tau_2=1.0$), enforcing $C_4$ rotational symmetry.
    \item \textbf{Configuration B (Anisotropic Target):} A rectangular lattice with $\tau = 1.2i$ ($\tau_2=1.2$). This value was selected based on the optimization trajectory in Sec. \ref{sec:experiments}-B, which indicated that increasing $\tau_2$ is the primary gradient direction for maximizing the objective $J(\tau)$.
\end{itemize}

\begin{figure}[htbp]
    \centering
    \includegraphics[width=1\linewidth]{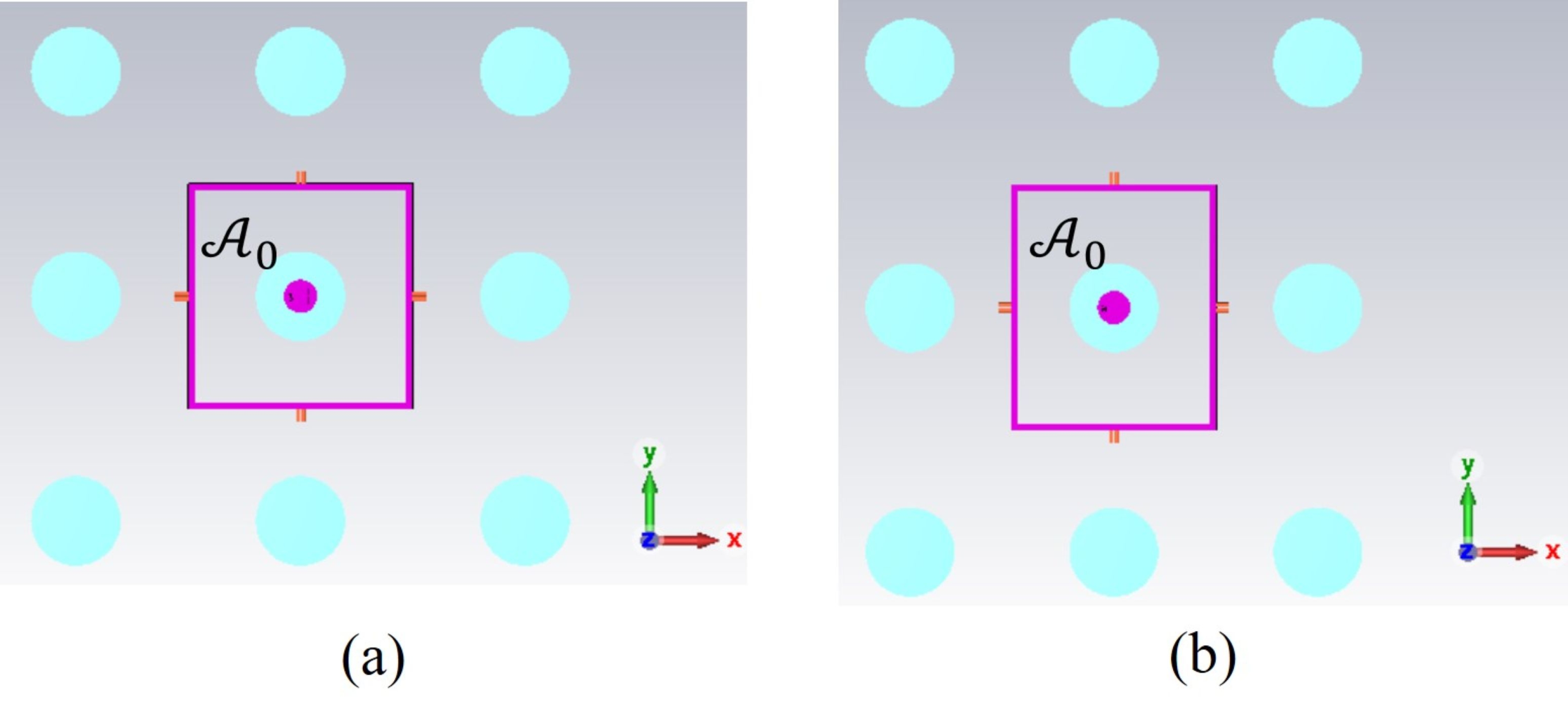}
    \caption{\textbf{Lattice Topology.} Top-down view of the dielectric metasurface unit cells. (a) Isotropic square lattice ($\tau_2=1.0$). (b) Anisotropic rectangular lattice ($\tau_2=1.2$), selected to verify the spatial dispersion splitting predicted by the gradient engine.}
    \label{fig:topology}
\end{figure}

Figure \ref{fig:topology} illustrates the topology of these two unit cells. The transformation from $\tau_2=1.0$ to $\tau_2=1.2$ breaks the four-fold rotational symmetry while preserving the unit cell area $\mathcal{A}_0$, consistent with the constraints applied in our derivative derivation.

\subsubsection{Verification of Angular Dispersion via Differential Contrast}

To rigorously quantify the engineered non-locality, we define the \textit{Anisotropy Contrast} $\Delta S_{ii}(\omega, \theta)$ as the magnitude difference between the scattering parameters of the two orthogonal polarization states:
\begin{equation}
    \Delta S_{ii}(\omega, \theta) = \bigl| \, |S_{ii}^{\phi=0}(\omega, \theta)| - |S_{ii}^{\phi=90}(\omega, \theta)| \, \bigr|, \quad i \in \{1, 2\}.
\end{equation}
Here, the indices $i=1$ and $i=2$ denote the fundamental TE ($E_y$-polarized) and TM ($E_x$-polarized) Floquet modes, respectively. This metric maps the "optimization distance" directly to the observable spectrum: a value of $\Delta \approx 0$ indicates effective isotropy (degenerate modes), while high contrast signifies strong polarization splitting.

\begin{figure}[htbp]
    \centering
    \includegraphics[width=1\linewidth]{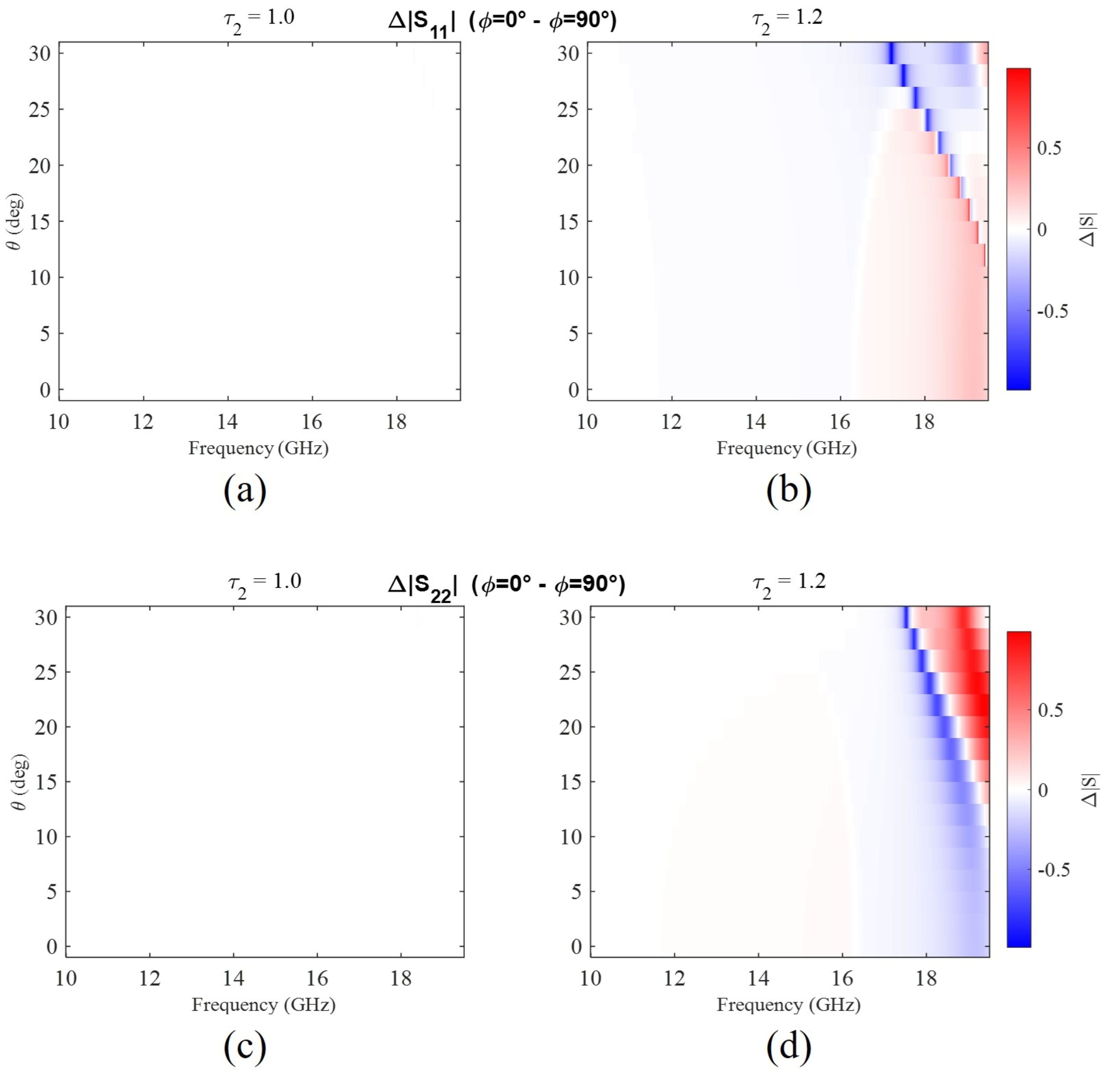} 
    \caption{\textbf{Full-Wave Verification of Symmetry Breaking.} The Anisotropy Contrast maps, $\Delta S_{ii} = ||S_{ii}^{\phi=0}| - |S_{ii}^{\phi=90}||$, plotted as a function of frequency and incidence angle. 
    \textbf{Left Column ($\tau_2=1.0$):} The dark fields in (a) $\Delta S_{11}$ and (c) $\Delta S_{22}$ indicate negligible polarization difference, confirming the symmetry-protected mode degeneracy of the square lattice.
    \textbf{Right Column ($\tau_2=1.2$):} The bright bands in (b) $\Delta S_{11}$ and (d) $\Delta S_{22}$ highlight the giant anisotropy engineered by the gradient optimizer. The mode splitting is substantial ($\sim 0.5$ GHz) and consistent across both ports, validating that the QMF-driven geometric perturbation $\tau \to 1.2i$ successfully unlocked the non-local spatial dispersion.}
    \label{fig:delta_map}
\end{figure}

Figure \ref{fig:delta_map} presents the full-wave simulation results arranged to compare the baseline square lattice ($\tau_2=1.0$) against the gradient-optimized rectangular lattice ($\tau_2=1.2$) across both port reflection parameters ($S_{11}$ and $S_{22}$).

For the isotropic baseline (Fig. \ref{fig:delta_map} (a) and (c)), the contrast maps are effectively featureless across the entire angular sweep ($\theta \in [0^\circ, 30^\circ]$). This confirms that the $C_4$ rotational symmetry preserves the spectral degeneracy of the TE-like and TM-like spatial dispersion surfaces, rendering the non-local response scalar-like.

In stark contrast, the optimized lattice (Fig. \ref{fig:delta_map}, (b) and (d)) reveals distinct, high-intensity bands in both $\Delta S_{11}$ and $\Delta S_{22}$. These bright regions correspond to the spectral bifurcation where the $x$- and $y$-polarized resonances split. Crucially, the consistency of this splitting pattern between the top ($S_{11}$) and bottom ($S_{22}$) ports confirms that the induced anisotropy is a robust bulk property of the lattice mode, rather than a superficial interface artifact. The optimization engine has successfully maximized the objective $J(\tau)$, physically manifesting as a "lifting" of the symmetry-protected degeneracy.

\section{Discussion and Limitations}
\label{sec:discussion}

The results presented in Section \ref{sec:experiments} demonstrate that the Quasi-Modular Form (QMF) gradient engine outperforms traditional finite-difference methods by orders of magnitude in both accuracy and speed. However, as an "exact" method rooted in analytic number theory, its performance characteristics are distinct from general-purpose numerical solvers. In this section, we critically analyze the computational bottlenecks, the domain of validity, and the potential for generalizing this framework to higher-order interactions.

\subsection{Computational Bottleneck: The Non-Holomorphic Sector}
The efficiency of our engine is strictly bifurcated between the \textit{reducible sector} (terms like $\tilde{\sigma}_2^{(4)}, \sigma_4^{(4)}$) and the \textit{non-reducible outlier} ($\sigma_4^{(2)}$).

For the reducible terms, the evaluation cost is effectively $O(1)$. The $SL(2, \mathbb{Z})$ reduction ensures rapid convergence of the Lambert series, requiring fewer than 10 terms to reach machine precision ($10^{-15}$) regardless of the lattice complexity.

In contrast, the term $\sigma_4^{(2)}$ remains the primary computational bottleneck. As it contains non-holomorphic Fourier components involving polylogarithms, it does not admit a polynomial reduction to $\{E_2, E_4, E_6\}$. Our hybrid strategy (Richardson extrapolation) improves the algebraic convergence from $O(K^{-2})$ to $O(K^{-4})$, but this is still asymptotically slower than the exponential convergence ($e^{-\pi \tau}$) of the modular forms.
In practice, for a typical lattice ($\tau_2 \approx 1.0$), evaluating $\sigma_4^{(2)}$ to $10^{-8}$ precision requires summing over a grid of $K \approx 20$, which is milliseconds compared to the microseconds for Eisenstein series. While this overhead is negligible for the gradient computation (since $\partial_\tau \sigma_4^{(2)}$ is analytic), it dominates the cost of the function evaluation $J(\tau)$.

\textit{Future Improvement:} For applications requiring extreme precision ($>10^{-12}$) for $\sigma_4^{(2)}$, the current direct summation could be replaced by Ewald summation or Poisson resummation techniques. However, these methods introduce special functions (incomplete Gamma functions) that complicate the derivation of exact analytic gradients with respect to $\tau$. The hybrid strategy chosen in this work represents a deliberate trade-off, prioritizing algebraic simplicity and code compactness over asymptotic optimality.

\subsection{Domain Validity and Geometric Constraints}
The mathematical derivation is valid for any $\tau \in \mathbb{H}$. However, numerical stability imposes a practical lower bound on the imaginary part $\tau_2 = \text{Im}(\tau)$.

As $\tau_2 \to 0$, the lattice becomes infinitely skewed, and the unit cell collapses. In this regime:
\begin{enumerate}
    \item \textbf{Physical Validity:} The Coupled Dipole Approximation (CDA) itself breaks down when the lattice spacing becomes comparable to the particle size ($L \sim 2r$), as near-field higher-order multipoles (e.g., octupoles) can no longer be neglected.
    \item \textbf{Numerical Cost:} For the Eisenstein series, the number of terms $N$ required for the error certificate scales as $1/\tau_2$ (after reduction to $\mathcal{F}$). More critically, for the lattice sum $\sigma_4^{(2)}$, the required grid size $K$ scales roughly as $\tau_2^{-1/2}$ to capture the slowly decaying tail along the flattened axis.
\end{enumerate}
In our implementation, we enforce a soft constraint $\tau_2 \ge 0.2$. This covers the entire physically relevant design space for dielectric metasurfaces (where typically $0.5 \le \tau_2 \le 2.0$). Designers working with extreme grazing-incidence structures or plasmonic chains may need to combine this method with 1D lattice sum accelerations (polylogarithms).

\subsection{Generalization to Higher-Order Multipoles}
While this work focused on the second-order spatial dispersion tensor $\bar{\mathcal{G}}^{(2)}$ (quadrupolar/dipolar-anisotropy order), the "Gradient Engine" methodology is universally applicable to any periodic interaction kernel that can be expanded in generalized Eisenstein series.

Higher-order nonlocal effects (e.g., $k^4$ terms or octupolar scatterers) involve lattice sums $S_n^{(m)}$ with higher decay orders $n=6, 8, \dots$.
\begin{itemize}
    \item \textbf{Reducibility:} Any term with $m \ge n$ (even) will map to a polynomial of derivatives of $G_n$. Since Ramanujan's identities close the ring for any weight, gradients for $S_6^{(8)}$ or $S_8^{(8)}$ can be derived automatically using the same recursive procedure used for $\tilde{\sigma}_2^{(4)}$.
    \item \textbf{Basis Extension:} The ring of modular forms is finitely generated by $E_4$ and $E_6$. Thus, no new special functions are needed for higher orders; only the polynomial complexity increases.
\end{itemize}
This suggests that the QMF framework can serve as a unified "Operating System" for the inverse design of complex periodic media, extending beyond electromagnetics to phononic crystals \cite{Sigmund2003} and fluid dynamic lattices, provided the governing Green's function admits a spectral expansion in cylindrical harmonics.

\subsection{Comparison with Complex-Step Differentiation}
A standard alternative to Finite Difference (FD) for obtaining high-precision derivatives is the Complex-Step Derivative Approximation (CSDA) \cite{Squire1998, Martins2003}, which avoids subtractive cancellation errors. However, we deliberately exclude CSDA as a baseline in this work for two fundamental reasons intrinsic to the modular framework.

First, the objective function $J(\tau)$ is natively defined on the complex upper half-plane $\mathbb{H}$. Implementing CSDA would require extending the domain from complex numbers to hyper-complex (or dual) numbers to rigorously distinguish the geometric imaginary part $\tau_2$ from the differentiation perturbation step $ih$. This adds unnecessary architectural complexity to the solver.

Second, and more critically, CSDA shares the same computational scaling as FD: it requires a full evaluation of the transcendental $q$-series for every design variable. In stark contrast, our QMF Analytic Engine obtains the gradient as a \textit{strictly algebraic byproduct} of the forward pass (as shown in Table \ref{tab:gradient_lookup}). By reusing the pre-computed Eisenstein values $\{E_2, E_4, E_6\}$, the marginal cost of computing the gradient is effectively zero. Therefore, the $O(h^4)$ Richardson extrapolation employed in Section \ref{sec:experiments} serves as a sufficient and rigorous independent validation, confirming that the analytic engine achieves the theoretical machine-precision floor without the overhead of complex-step evaluations.

\section{Conclusion}
\label{sec:conclusion}

In this work, we  present a rigorous analytical framework for the inverse design of spatially dispersive metasurfaces, addressing the fundamental "accuracy-stability" trade-off inherent in traditional numerical optimization. By bridging computational electromagnetics with the number-theoretic properties of Quasi-Modular Forms (QMF), we constructed an end-to-end \textit{Analytic Gradient Engine} for the periodic Dyadic Green's Function.

Our methodology replaces the approximate finite-difference calculus with exact algebraic operations. Through the application of Ramanujan's differential identities, we derived closed-form expressions for the sensitivity of the interaction matrix with respect to the lattice geometry $\tau$. Crucially, we address the practical computational challenges by:
\begin{enumerate}
    \item \textbf{Exact Reduction:} Mapping the physical lattice sums to a minimal basis of Eisenstein series and handling the non-reducible outlier $\sigma_4^{(2)}$ via a robust hybrid strategy.
    \item \textbf{Numerical Stability:} Implementing an $SL(2, \mathbb{Z})$ domain reduction algorithm with automatic error certificates, ensuring machine-precision accuracy ($10^{-15}$) even for highly skewed lattices.
    \item \textbf{Geometric Consistency:} Deriving a fixed-area chain rule that allows the optimizer to explore the modular manifold while respecting physical scaling constraints.
\end{enumerate}

Experimental validation confirmed the superiority of this approach. The analytic engine demonstrated a $6.5\times$ speedup over finite-difference baselines and successfully navigated the complex optimization landscape to maximize non-local anisotropy. Full-wave simulations of the resulting designs validated the physical predictions, observing a distinct splitting of spatial dispersion modes in the optimized lattice.

This framework opens new avenues for the design of complex periodic media. While this study focused on 2D Bravais lattices and quadrupolar interactions, the underlying algebraic structure generalizes to higher-order multipoles and can be extended to 3D lattices via the theory of Epstein zeta functions \cite{Epstein1903, Buchheit2024}. Furthermore, the ability to compute exact derivatives near spectral singularities makes this method particularly well-suited for the emerging field of topological photonics \cite{Lu2014, Ozawa2019}, where the sensitivity of Berry curvature and topological invariants to geometric perturbations is of paramount importance.

\section*{Acknowledgments}
We thank Yajun An (University of Washington, Tacoma) for her careful review and insightful suggestions on the manuscript.

\appendix
\section{Standard Eisenstein Series Definitions}
For reference, the standard Eisenstein series $E_{2k}(\tau)$ used in the reduction formulas are defined as:
\begin{equation}
    E_{2k}(\tau) = 1 + \frac{2}{\zeta(1-2k)} \sum_{n=1}^{\infty} \frac{n^{2k-1} q^n}{1-q^n}, \quad q = e^{2\pi i \tau}.
\end{equation}
Their connection to the lattice sums $G_{2k}(\tau) = \sum_{(m,n)\ne(0,0)} (m+n\tau)^{-2k}$ is given by $G_{2k} = 2\zeta(2k)E_{2k}$.
\section*{Code and Data Availability}

The data supporting the findings of this study are generated using the analytic algorithms and numerical procedures fully described in Sections \ref{sec:reduction}--\ref{sec:numerics} of this article. No experimental datasets were used. The Python implementation of the Quasi-Modular Form gradient engine and the optimization scripts used to generate the results are available from the corresponding author upon reasonable request.

\bibliographystyle{elsarticle-num}
\bibliography{references}

\end{document}